\documentclass[fleqn,usenatbib, article]{mnras}
\usepackage{amsmath}
\usepackage{mathptmx}
\usepackage{txfonts}

\usepackage[T1]{fontenc}
\usepackage{ae,aecompl}
\usepackage{dblfloatfix}

\usepackage{graphicx}
\usepackage[pdf]{pstricks}
\usepackage{epstopdf}
\usepackage{amssymb}
\usepackage{latexsym}
\usepackage{tikz-cd}
\usepackage{bm}
\usepackage{natbib,twoopt}

\newcommand{\bam}{\texttt{BAM} }
\newcommand{\pat}{\texttt{PATCHY} }

\title[Accuracy of bias mapping techniques]{The bias of dark matter tracers: assessing the accuracy of mapping techniques }

\author[Pellejero-Iba{\~n}ez et al.]{\parbox{\textwidth}
  {M.~Pellejero-Iba{\~n}ez \thanks{mpellejero@dipc.org}$^{1}$,
  A. Balaguera-Antol\'{\i}nez\thanks{balaguera@iac.es}$^{2,3}$,
  Francisco-Shu Kitaura\thanks{fkitaura@iac.es}$^{2,3}$} \and Ra\'ul E. Angulo$^{1,4}$, Gustavo Yepes$^{5,6}$, Chia-Hsun Chuang$^{7}$, Guillermo Reyes-Peraza$^8$, \and Mathieu Autefage$^9$, Mohammadjavad Vakili$^{10}$ \& Cheng Zhao$^{11}$
  \\ \\
  $^{1}$Donostia International Physics Centre   (DIPC), Paseo Manuel de Lardizabal 4, 20018 Donostia-San Sebastian, Spain\\
  $^{2}$Instituto de Astrof\'{\i}sica de Canarias, s/n, E-38205, La Laguna, Tenerife, Spain\\
  $^{3}$Departamento de Astrof\'{\i}sica, Universidad de La Laguna, E-38206, La Laguna, Tenerife, Spain\\
  $^{4}$IKERBASQUE, Basque Foundation for Science, 48013, Bilbao, Spain.\\
  $^{5}$Departamento de F\'{\i}sica Te\'orica, Universidad Aut\'onoma de Madrid, Cantoblanco E-28049, Madrid, Spain \\ 
  $^{6}$Centro de Investigaci\'on Avanzada en F\'{\i}sica Fundamental,  Facultad de Ciencias, Universidad Aut\'onoma de Madrid, E-28049 Madrid, Spain \\
  $^{7}$Kavli Institute for Particle Astrophysics and Cosmology, Stanford University, 452 Lomita Mall, Stanford, CA 94305, USA\\
  $^{8}$Instituto de F\'{\i}sica Te\'orica, (UAM/CSIC), Universidad Aut\'onoma de Madrid, Cantoblanco, E-28049 Madrid, Spain \\
  $^{9}$T\'el\'ecom Physique Strasbourg P\^ole API, 300 Bd S\'ebastien Brant, 67400 Illkirch-Graffenstaden, France  \\
  $^{10}$Leiden Observatory, Leiden University, Niels Bohrweg 2, NL-2333 CA Leiden, The Netherlands \\
  $^{11}$\'Ecole polytechnique f\'ed\'erale de Lausanne, Route Cantonale, 1015 Lausanne, Switzerland
 }
\begin{document}
\label{firstpage}
\pagerange{\pageref{firstpage}--\pageref{lastpage}} 
\maketitle

\begin{abstract}
  We present a comparison between approximated methods for the construction of mock catalogs based on the halo-bias mapping technique. To this end, we use as reference a high resolution $N$-body simulation of 3840$^3$ dark matter particles on a 400$h^{-1}\rm{Mpc}$  cube box from the Multidark suite.
  In particular, we explore parametric versus non-parametric bias mapping approaches and compare them at reproducing the halo distribution in terms of the two and three point statistics down to $\sim 10^8\,{\rm M}_{\odot}\,h^{-1}$ halo masses.
  Our findings demonstrate that the parametric approach remains inaccurate even including complex deterministic and stochastic components. On the contrary, the non-parametric one is indistinguishable from the reference $N$-body calculation in the power-spectrum beyond $k=1\,h\,{\rm Mpc}^{-1}$,  and in the bispectrum  for typical configurations relevant to baryon acoustic oscillation analysis. 
  We conclude, that approaches which extract the full bias information from $N$-body simulations in a non-parametric fashion are ready for the analysis of the new generation of large scale structure surveys.
  
\end{abstract}
\begin{keywords}
cosmology: -- theory - large-scale structure of Universe 
\vspace{-0.7cm}
\end{keywords}
\section{Introduction}\label{sect1}
With the starting-gun for some of the most impressive observational campaigns in large scale structure about to be shot (e.g. Euclid, \citealt[][]{Euclid}, and DESI, \citealt[][]{DESI}), the quest for precise and accurate covariance matrices, aimed at assessing the uncertainties in the measurements of cosmological observables such as redshift space distortions \cite[e.g.][]{1987MNRAS.227....1K} and baryonic acoustic oscillations \cite[e.g.][]{1998ApJ...496..605E} has become a task of front line research \cite[see e.g.][]{2013PhRvD..88f3537D,2013MNRAS.432.1928T,2015MNRAS.454.4326P}. The construction of large sets of catalogs based on $N$-body simulations has been adopted as the standard path to obtain robust estimates on the errors of cosmological observables. Such approach is nevertheless unpractical, given the considerable amount of time and/or memory requirements that a state-of-the-art $N$-body simulation requires to generate hundreds to thousands of realizations with the desired cosmological volumes and mass resolution. Several approaches have been suggested in the literature to speed-up the construction mock catalogs 
\citep[e.g.][]{1996ApJS..103....1B,2002MNRAS.329..629S,2002MNRAS.331..587M,2013JCAP...06..036T, 2016MNRAS.459.2118K,2016MNRAS.459.2327I,2016MNRAS.463.2273F,2003ApJ...593....1B, 2005ApJ...630....1Z,2013MNRAS.428.1036M,2015MNRAS.447..437M,2014MNRAS.437.2594W,2015MNRAS.450.1856A,2015MNRAS.446.2621C,Angulo:2013gya}. The idea in some of these approaches (such as \pat \citealt{2014MNRAS.439L..21K}) is to rely on the smooth large-scale dark matter field obtained from approximate gravity solvers, and populate it with halos (or galaxies) following some bias prescriptions, in what is known as the \emph{bias mapping technique}. The precision of this type of approach in the resulting halo catalogs is only acceptable, according to scientific requirements of forthcoming surveys, until intermediate scales ($k\sim 0.2-0.3\, h\,{\rm{Mpc}}^{-1}$ in Fourier space), \citealp[see e.g.][]{2016MNRAS.456.4156K,10.1093/mnras/stz507,10.1093/mnras/sty2964, 10.1093/mnras/sty2757}. Higher precision ($\sim 1-2\%$) towards smaller scales beyond $k=1\, h\,{\rm{Mpc}}^{-1}$ (including the Nyquist frequency) has been only reached by the recently published \emph{Bias Assignment Method} (\texttt{BAM} hereafter, \citealt{2019arXiv190606109B,2019MNRAS.483L..58B}).
\begin{figure*}
  \includegraphics[width=5.8cm]{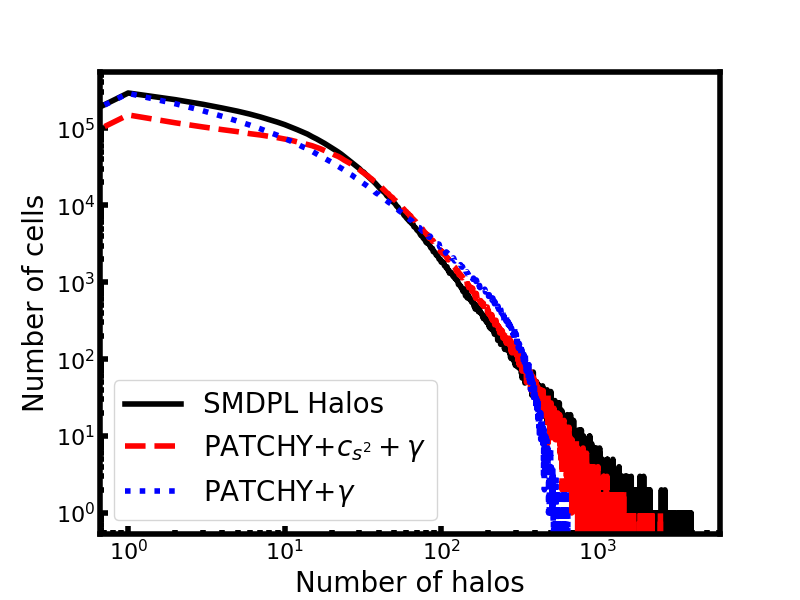}
  \includegraphics[width=5.8cm]{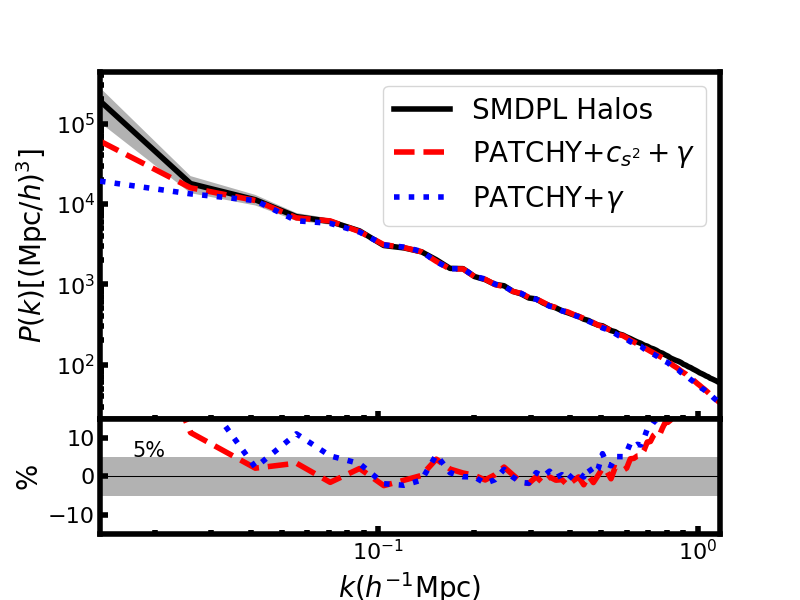}
  \includegraphics[width=5.8cm]{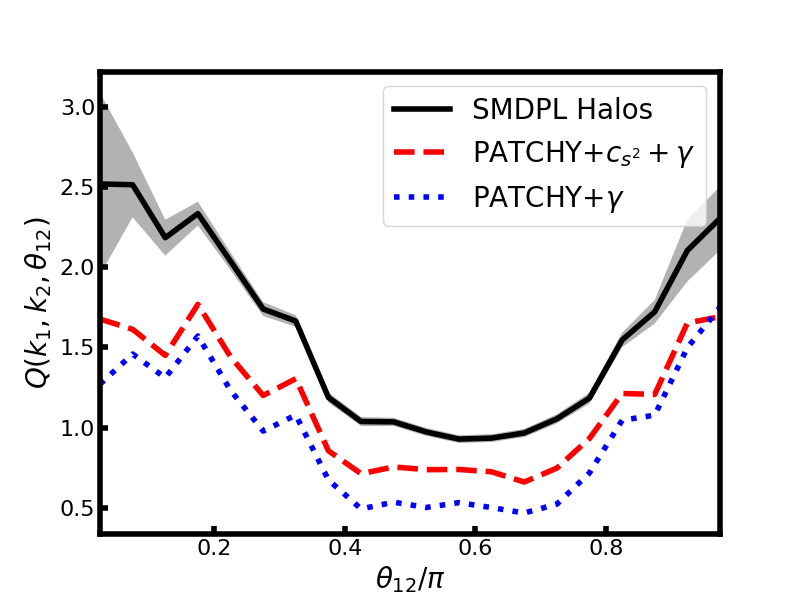}\\
  \includegraphics[width=5.8cm]{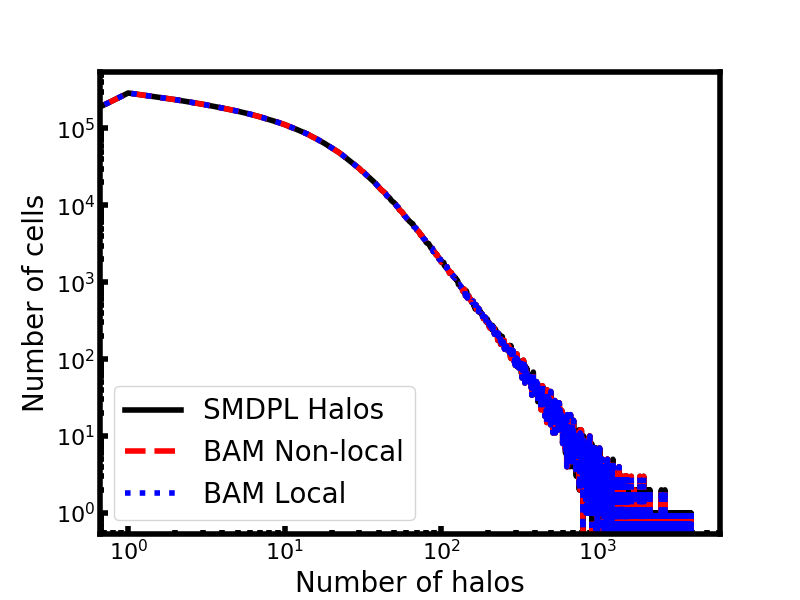}
  \includegraphics[width=5.8cm]{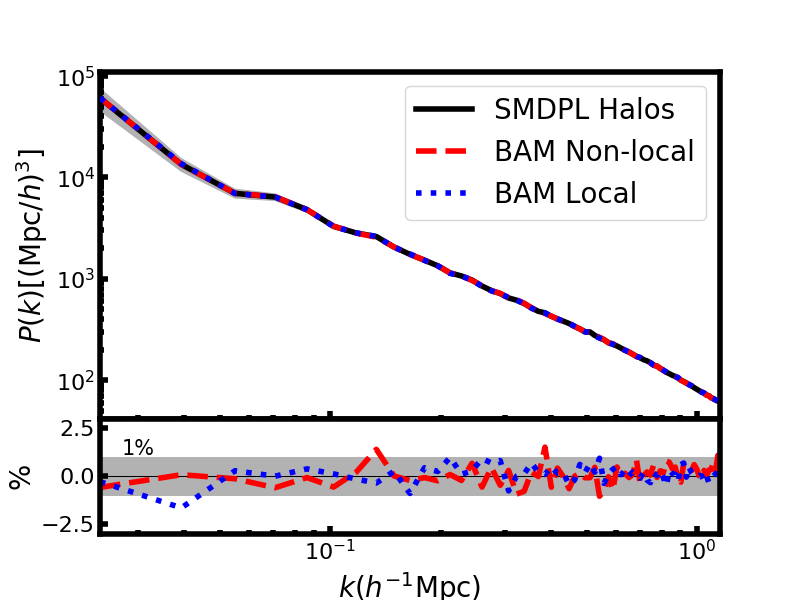}
  \includegraphics[width=5.8cm]{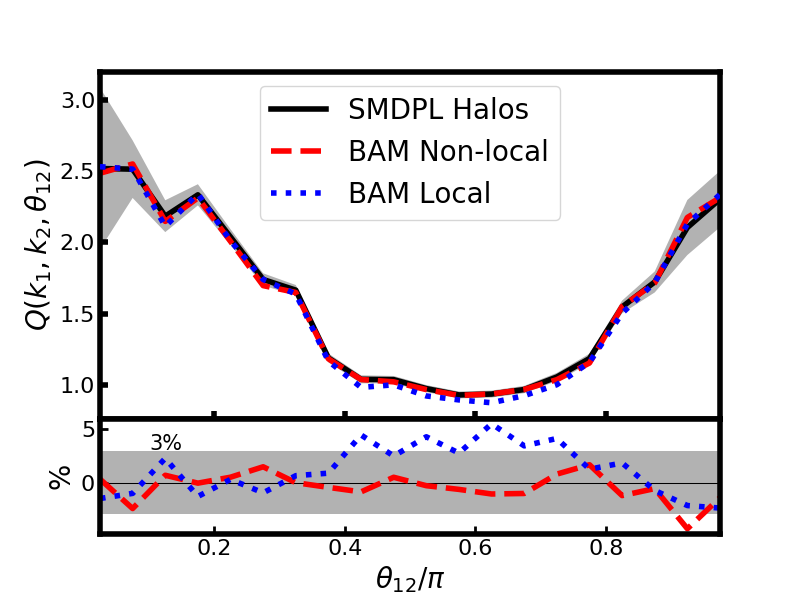}
\caption{Upper row: comparison of the probability density functions (PDF, left), power spectra ($P(k)$, middle) and bispectra ($Q(k_1,k_2,\theta_{12})$, right) obtained from the sampling of a ADMF at $z=0$ using Eq.~(\ref{eq:deterministic_nonlocal}) with $c_{s^{2}}=0$ (local) and $c_{s^{2}}\neq 0$ (non-local). 
Lower row: PDF (left panel), power spectra ($P(k)$, middle panel) and bispectrum ($Q(k_1,k_2,\theta_{12})$, right panel) obtained with \bam. The configuration of the bispectra is $k_2=2k_1=0.2\, h \,{\rm{Mpc}}^{-1}$. The subsets in the panels show the percentage differences between the reference and approximated outputs.} 
\label{Pk_patchy_OldBias}
\end{figure*}

The goal of this paper is to assess whether bias mapping methods such as \pat and \texttt{BAM}, already successfully applied to mass-scales of the order of $\sim 10^{12}\, {\rm M}_{\odot}h^{-1}\,$ (see e.g. \citealt{2017MNRAS.472.4144V} and \citealt{2019MNRAS.483L..58B}), are still capable to capture the main features of halo bias when applied to a high resolution (low mass-scales) $N$-body simulation. The \pat method has been shown to reach high level of accuracy circumventing the limitations of the approximate gravity solvers \cite[see e.g.][]{2015MNRAS.452..686C, 2015MNRAS.450.1836K,  2017MNRAS.472.4144V}, being able to make large amount of precise mock galaxy catalogues \citep[e.g.][]{2016MNRAS.456.4156K} with very low memory and computational requirements, while \texttt{BAM} has been shown to replicate to percent precision summary statistics of halo catalogs (with approximately same computational time and memory requirements as \pat). It is therefore timely to assess whether these methods can be still applied to the expected galaxy catalogs from Euclid or DESI, probing lower halo mass ranges, or if a change towards non-parametric approaches forecasts better results.

In order to perform this comparison, we use the {\sc Small MultiDark Planck} simulation\footnote{See \url{http://dx.doi.org/10.17876/cosmosim/smdpl/} for details about the \texttt{SMDPL}}  (\texttt{SMDPL} hereafter), which belongs to the series of {\sc MultiDark} simulations \citep{Klypin:2014kpa} with Planck cosmology \citep{planck2015-xiii}. The simulation consists in a set of $3840^3$ particles dark matter particles in a comoving volume of $V_{s}=(400$ $h^{-1}$Mpc$)^{3}$. Dark matter halos and subhalos are identified with the \texttt{Rockstar} algorithm \citep[][]{2013ApJ...762..109B}, with a minimum tracer mass of $\sim 2\times10^8 {\rm M}_{\odot} h^{-1}$ at $z=0$ ($\sim 8.2\times 10^{7}$ number of halos and subhalos). Although in what follows we use the complete sample of halos and subhalos as tracers of the dark matter field, we will refer to them generically as ``halos'' or ``tracers''. Note that this is the first time that both techniques are tested against a sample of halos and subhalos.

Both the \texttt{PATCHY} method and \texttt{BAM} are based on the concept of stochastic bias \citep[e.g.][]{1999ApJ...520...24D} in order to generate a halo density field from an approximated dark matter density field (ADMF hereafter). This bias is expressed as a conditional probability distribution (CPD hereafter), accounting for the probability of obtaining a given number of halos conditional to local and non local properties of the ADMF. In \pat the CPD is characterized by a mean and a scatter around that mean, both expressed through a number of parameters describing a given model for the mean and stochasticity. In \texttt{BAM} instead, the CPD is \emph{directly measured} from the reference simulation in combination with the ADMF. For both methods, the ADMF is obtained by the evolution of downgraded initial conditions from the \texttt{SMDPL} simulations. In our current set-up, we performed this operation starting with a white noise embedded in a $3840^3$ mesh into one of $160^3$ grid cells. These downgraded initial conditions were thereafter evolved until redshift $z=0$ using approximated gravity solvers such as \texttt{FastPM} \citep{2016MNRAS.463.2273F} or the Augmented Lagrangian Perturbation Theory (ALPT) \citep{Kitaura:2012tj} on an equally resolved mesh ($160^3$). While for the \texttt{BAM} approach the choice of gravity solver is not relevant, that is not the case for \pat, in which the signal of the power spectrum at intermediate scales can affect the performance of the method. We therefore adopt \texttt{FastPM} on a mesh of $160^3$ cells to evolve the downgraded initial conditions up to redshift $z=0$. Although not specifically shown in this study, a different resolution of $192^3$ was used and similar results were achieved. 

The outline of this paper is as follows. We describe the parametric bias prescription in \S ~\ref{sec:patchy}, and show the results of the fits of such prescription to the \texttt{SMDPL} simulation. We show the results from the non-parametric approach in \S~\ref{sec:bam}. We discuss possible signatures of non local bias in \S~\ref{sec:nl} and end-up with conclusions and discussion.

\section{Parametric bias prescription: \pat}\label{sec:patchy}
In \pat \citep{2014MNRAS.439L..21K}, the deterministic component  (or mean of the CPD) relating the halo ($\rho_{\rm{h}}$) and the dark matter ($\rho_{\rm{m}}$) density fields is written in terms of the matter overdensity $\delta \equiv \rho/\bar{\rho}-1$ as \citep[see e.g.][]{Neyrinck:2013ezr,2014MNRAS.439L..21K}
\begin{eqnarray}
\label{eq:deterministic_nonlocal}
\langle (1+\delta_{\rm h}) \rangle & \equiv & \langle  (1+\delta_{\rm h})| (1+\delta_{\rm m}) \rangle \\ 
& = & \big((1+\delta_{\rm  m})^\alpha + c_{\rm s^{2}}s^{2}\big) \,
 \theta\big(\delta_{\rm m} - \delta_{\rm th}\big) \, \exp \big[-\big((1+\delta_{\rm m})/\rho_{\epsilon}\big)^{\epsilon}\big]\,. \nonumber
 \end{eqnarray}
In this expression $\delta_{\rm th}$ is an overdensity threshold, aimed at suppressing the formation of halos in under-dense regions. The threshold bias \citep[e.g.][]{Kaiser:1984sw,Bardeen:1985tr,Sheth:1999su} is represented by a step function, $\theta \big(\delta_{\rm m} - \delta_{\rm th}\big)$, and an exponential cutoff, $\exp \big[-\big((1+\delta)/\rho_{\epsilon}\big)^{\epsilon}\big]$. On the other hand, the term $(1+\delta_{\rm m})^\alpha$ accounts for nonlinear behaviour.
In Eq.~(\ref{eq:deterministic_nonlocal}) we have introduced a second order non-local term proportional to $s^{2}=s_{ij}s^{ij}$ \citep[see e.g.][]{2009JCAP...08..020M,2018PhR...733....1D,Abidi:2018eyd}, where $s_{ij}\equiv  \partial_{ij}\Phi-\delta^{\rm K}_{ij}\delta_{\rm m}/3$ ($\delta^{\rm K}_{ij}$ stands for the Kronecker delta) are the elements of the traceless tidal field tensor \citep[see e.g.][]{2007MNRAS.375..489H}, written in terms of the gravitational potential $\Phi$ obtained from the Poisson equation $\nabla^{2}\Phi=\delta_{\rm m}$. 

The stochasticity around this deterministic bias is modelled by a negative binomial distribution (\citealt{2014MNRAS.439L..21K, 2017MNRAS.472.4144V}), with  expected number of halos in each cell (given by the deterministic bias component) $\lambda_{\rm h} \equiv \langle N_{\rm{h}}\rangle_{{\rm d} V} = f_N \langle 1+\delta_{\rm h} \rangle_{{\rm d}V} {\rm d}V$ and an scatter characterized by a Poisson distribution with a parameter $\beta$ allowing for deviations from the latter. Here, $f_N \equiv \bar{N}/\langle 1+\delta_h \rangle_V$ controls the halo number density (i.e. it is fixed by the number density of the reference catalog). Hence, the probability of having $N_{\rm h}$ objects in a volume element with mean $\lambda_{h}$ is given by
\begin{equation}
P(N_{\rm h}|\lambda_{\rm h}, \beta) = \left(\frac{\lambda_{\rm h}^{N_{\rm h}}}{N_{\rm h}!}\, 
\right) \left(\frac{\Gamma(\beta+N_{\rm h})}{\Gamma(\beta)(\beta + \lambda_{\rm h})^{N_{\rm h}}}\right) 
\left(1+\frac{\lambda_{\rm h}}{\beta}\right)^{-\beta},
\label{eq:devpois}
\end{equation}
where $\Gamma(\beta)$ is the Gamma function \footnote{For $\beta\rightarrow\infty$ it can can be shown that the second term in Eq.~(\ref{eq:devpois}) goes to one, reducing the full expression to a Poisson distribution.}. We further test the addition of an extra dependency over $\lambda_{\rm h}$ by substituting $\beta \rightarrow \beta (\lambda_{\rm h}/\bar{N})^\gamma$, where $\bar{N}$ is the mean number of halo in the volume. \pat uses Eq.~(\ref{eq:devpois}) to sample a halo density field (number counts) according the dark matter density (local) and the tidal field (non local), constrained to generate a sample with the mean number density of the reference simulation. From this sampled halo density field we can measure the summary statistics (i.e, one, two point statistics etc).

We have constrained the set of parameters defining the bias models (Eqs.~(\ref{eq:deterministic_nonlocal}) and (\ref{eq:devpois})), $\{\delta_{\rm th}, \alpha, \rho_{\epsilon}, \epsilon, \beta, \gamma, c_{s^{2}}\}$ following the procedures of \citet[][]{2015MNRAS.450.1836K} and \citet{2017MNRAS.472.4144V}, based on Markov-Chain likelihood analysis of the parameter space. We used the probability density function (PDF) and the power spectrum (up to $k=0.4h/{\rm{Mpc}}$) as observables, and assumed independent likelihoods with a Poisson variance for the PDF ($\sigma^2_{\rm PDF}=N_n$, where $N_n$ is the number of cells containing $n$ dark matter halos), and a Gaussian variance for the power spectrum ($\sigma^2_P = 4\pi^2P_{\rm ref}(k)/(V_{s}k^2\Delta k )$, where $P_{\rm ref}(k)$ denotes the power spectrum measured from the reference simulation). 
We summarise the constraints on the bias parameters in Table \ref{table:BestFitMock}.

\begin{figure*}
  \includegraphics[width=18cm, height=7cm]{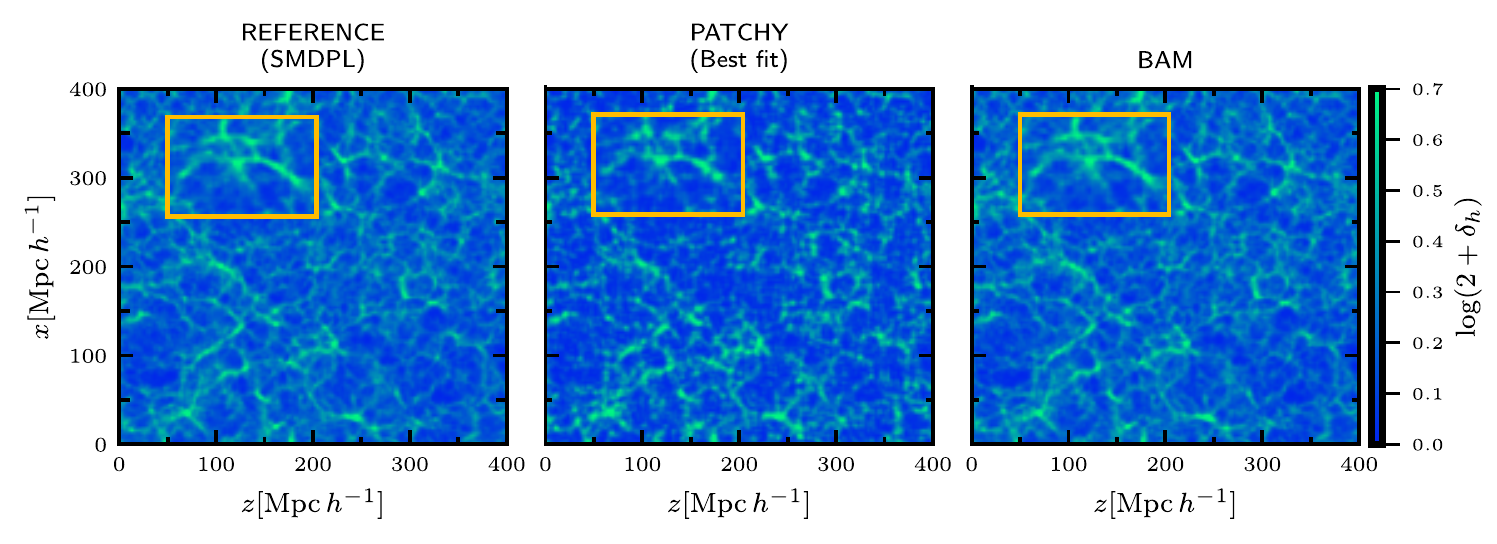}
\caption{Slices of $25$ Mpc$\,h^{-1}$ thickness of the halos overdensity field from the reference \texttt{SMDPL} simulation (left), \pat (center with the best fit parameters in Eq.~(\ref{eq:deterministic_nonlocal}) and (\ref{eq:devpois}) (with a non-zero value of $c_s^2$) and \texttt{BAM} (right). The yellow square zooms a specific place of the cosmic web of size $50\times100({\rm{Mpc}}/h)^2$. The structures obtained from the \pat approach display a slightly less evident filamentary structure when compared to the reference.} 
  \label{densf}
 \end{figure*}

\begin{figure*}
  \includegraphics[width=18cm, height=6.5cm]{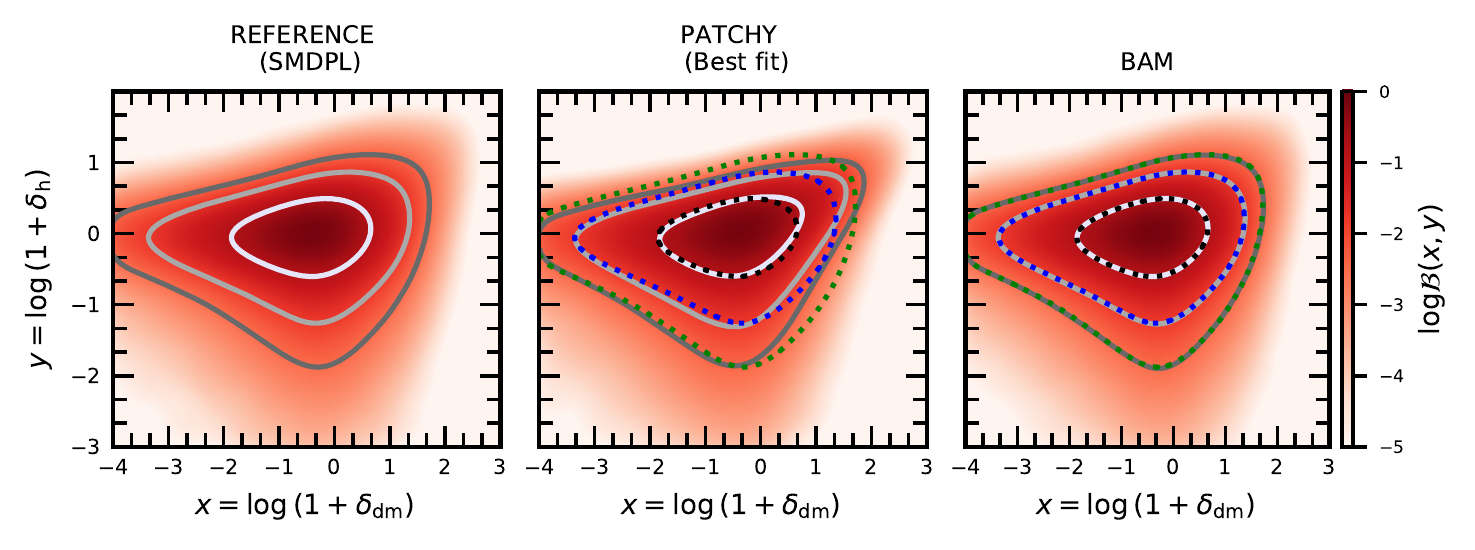}
\caption{Two-dimensional projection of the halo bias (normalized to its maximum value). In each plot, the solid lines encloses the region containing $68\%, 95\%$ and $99\%$ of the total number of cells ($160^{3}$). In order to facilitate the comparison, the dotted contours in the second and third panel denotes those form the reference (first panel). For this plot the halo overdensity and the dark matter overdensity were computing using a CIC interpolation scheme.} 
  \label{biasF}
 \end{figure*}

\begin{table}
\centering
\begin{tabular}{c c c}
       \hline
	    \textcolor{blue}{Parameter}	&	\textcolor{blue}{Local} &	\textcolor{blue}{Non-local}  	\\
     \hline 
$	\alpha          	$&$	1.37 \pm 0.02	 $&$  0.72 \pm 0.03 $\\
$	\rho_{\epsilon} 	$&$ 1.94 \pm 0.08  $&$  0.14 \pm 0.06 $\\
$	\epsilon            $&$	0.4 \pm 0.2	 $&$ -1.4 \pm 0.5 $\\
$	\beta            	$&$ 5.9 \pm 0.2 $&$ 8.0 \pm 0.2 $\\
$	\gamma            	$&$ 0.38 \pm 0.09 $&$ 0.04^{+0.09}_{-0.04}$\\
$	c_{s^2}            $&$ \rm{-} $&$(5 \pm 3)\times 10^{-4}
$\\
     \hline
 \end{tabular}
\caption{Mean values with $68\%$ errors of the parameters characterizing the model of halo-bias in Eqs.~(\ref{eq:deterministic_nonlocal}) and (\ref{eq:devpois}) in \pat. These values are obtained from the Markov-Chain Monte Carlo analysis using as observables the PDF and the power spectrum. The threshold density is fully consistent with the value $\delta_{\rm th}\sim -1$.}
\label{table:BestFitMock}
\end{table}

The resulting PDF and power spectrum are shown in top panel of Fig.~\ref{Pk_patchy_OldBias}. 
The PDF is not well fitted by any of the cases under study. \pat without non-local terms is able to fit the \texttt{SMDPL} number of cells for those with small number of halos but fails when increasing the number of halos per cell. If the non-local term is included and we still force a correct fit for the $P(k)$, the PDF is correctly fitted for cells with large number of halos but fails in the low number end. Focusing now on the $P(k)$, the local model reaches an accuracy of $\sim5\%$ up to scales of $\sim 0.55\,h\,{\rm{Mpc}}^{-1}$, while the non-local extension remains within the same accuracy up to scales of $\sim 0.65\, h\,{\rm{Mpc}}^{-1}$. Note again that for this result we used the complete halo sample of $M>2\times10^8 {\rm M}_{\odot} h^{-1}$  (which roughly corresponds to a linear bias of $\sim 0.65$).

Although not explicitly shown in Fig. \ref{Pk_patchy_OldBias}, the inclusion of the parameter $\gamma$ contributes to a marginal improvement in accuracy of the PDF and $P(k)$ when only accounting for local terms (thus its non-zero value of Table \ref{table:BestFitMock}). Nonetheless, when accounting for the non-local term $s^2$, the best-fit value of $\gamma$ is consistent with zero (see Table \ref{table:BestFitMock}). The value of $\epsilon$ changes sign when including non-local terms, changing the behaviour of the exponential cut-off. Furthermore, when including non-local terms, a larger $\beta$ is favoured indicating closer resemblance with Poissonity. The value of $\delta_{\rm{th}}$ is consistent with $-1$ indicating that, for the current mass-limit, the step function of Eq. (\ref{eq:deterministic_nonlocal}) does not suppress the formation of halos in under-dense regions. Therefore, the full dark matter simulation is preferred over density cuts over it. In any case, we stress that neither the one- nor the two-point statistics are reproduced to the desired per-cent precision by the model with best-fit values, indicating that such approach cannot account for a realistic description of the halo-bias at such low mass scales.

The third panel of Fig.~\ref{Pk_patchy_OldBias} shows the signal of the bispectrum obtained from sampling the ADMF with Eq.~(\ref{eq:deterministic_nonlocal}), using the best fit parameters. The estimates are biased with respect to the reference, regardless the presence of the non-local terms in the bias model. In order to provide an estimate of the error-bars for the bispectrum, we use the expression \citep[see e.g.][]{fry_article,1998ApJ...496..586S,Scoccimarro:2000sn,2015JCAP...10..039A}
\begin{equation}\label{err_bis}
\sigma_B^2 \approx s_B \frac{k_f}{V_B}P(k_1)P(k_2)P(k_3) \, , 
\end{equation}
where $s_B = 6, 2, 1$ for equilateral, isosceles and scalene configurations in Fourier space, $V_B\approx 8\pi^2k_1k_2k_3\Delta k$ and the fundamental mode is $k_f = 2\pi/L_{\rm{box}}$. We include this variance to have a taste of how much is the \pat methodology biased in terms of the cosmic variance. Using this estimates for the variance, we quantify the difference between the bispectrum generated by \pat and the reference to $\sim70\%$ ($\sim 10\sigma_B$). We therefore conclude that the model of Eq.~(\ref{eq:deterministic_nonlocal}) cannot describe the higher order statistics of a halo sample with the mass resolution of the \texttt{SMDPL}. 

In the next section, when mentioning statistical significance, we will be referring to this expression, though will not explicitly present the error bars in order to avoid clutter.

\begin{figure}
    \hspace{-1.1cm}
    \includegraphics[width=13cm]{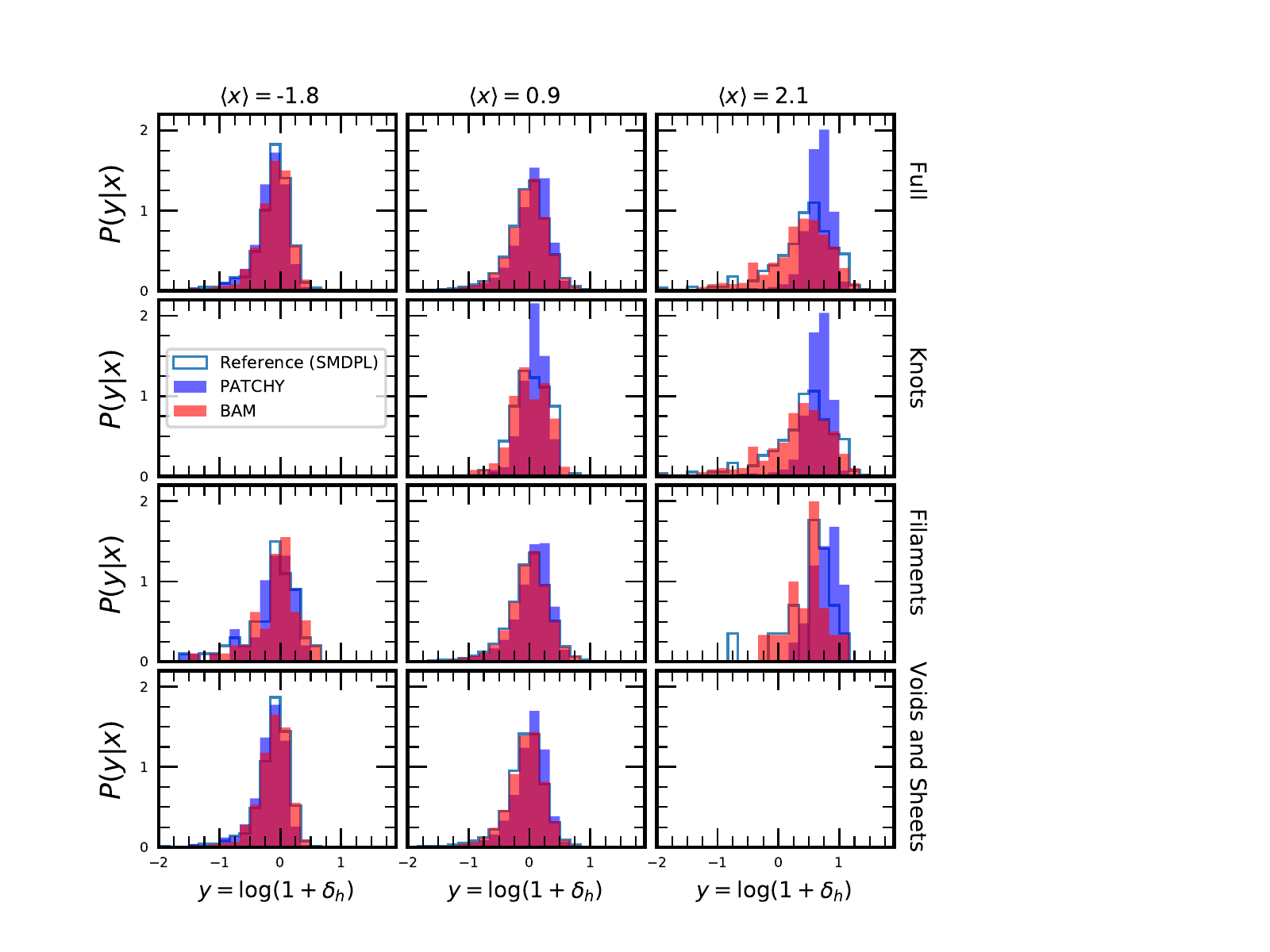}
    \vspace{-0.8cm}
    \caption{Examples of halo-bias, expressed as the probability distribution $P(y|x)$ of halo overdensity $y\equiv \log (1+\delta_{h})$ in different bins of ADM densities (expressed as $x\equiv \log (1+\delta_{\rm m})$) and for the full samples (top-row) and different cosmic-web types (second to fourth row). In all cases, the discrepancies between the reference and \pat are mildly evident on the low dark matter density bins. For high values of dark matter density, \pat displays a significant lack of cells with low values of halo overdensities, as can be visually inferred from Fig.\ref{densf}. Knots (voids and filaments) are not found on low (high) density regions shown in this plot. Note that the bias from \texttt{BAM} and the reference are statistically indistinguishable.}
  \label{bias_pdf}
\end{figure}

\section{Non parametric approach: the Bias Assignment Method}\label{sec:bam}
\bam \citep{2019arXiv190606109B,2019MNRAS.483L..58B} is an example of a non-parametric approach to the mapping of halo bias \citep[for other non parametric approaches based on machine learning techniques see also][]{2017arXiv170602390M,2018arXiv180804728M,2018arXiv181106533H,McClintock:2019sfj}. The method statistically maps the halo bias (i.e, the CPD) measured from the reference simulation and the ADMF (a \texttt{FastPM} generated density field in this case) into the latter, producing a halo sample with the PDF of the reference. This is preformed within an iterative process, in which the ADMF is convolved with a kernel computed from the ratio between the halo power spectrum of the reference simulation and the power spectrum of the halo field produced in the previous iteration. The \bam kernel is subject to a Metropolis-Hasting selection criteria, which guarantees that within each iteration the power spectrum of the new halo number counts approaches that of the reference. Indeed, after some iterations (for the current set-up, typically $\sim 10$), the halo field produced by \bam is such that its power spectrum differs $\sim 1\%$ with respect to that from the reference\footnote{For more details on the iterative process see figure 2 from \cite{2019arXiv190606109B}}. In \bam the halo-bias is measured as a function of a set of properties of the underlying ADMF such as the local density. We also use \emph{non-local contributions} such as the \emph{cosmic web-type} (i.e, knots, filaments, sheets and voids), classified using the eigenvalues of the tidal field \citep[see e.g.][]{2007MNRAS.375..489H}. Another non-local quantity used in \bam is the mass of collapsing regions. These regions are defined as sets of friend-of-friends cells (classified as knots) \citep[e.g.][]{2015MNRAS.451.4266Z}.

The bottom panels from Fig.~\ref{Pk_patchy_OldBias} show summary statistics obtained from the non-parametric method. In the first and second column we show the PDF and power spectrum, which, as exposed before, display percent differences by construction (regardless whether non-local properties are also included). More relevant is the information shown in the third, where we display the reduced bispectrum measured at the configuration $k_1=0.1\,h\,{\rm{Mpc}^{-1}}$ and $k_2=0.2\,h\,{\rm{Mpc}}^{-1}$. It is impressive how \texttt{BAM}  reproduces (at the resolution of the \texttt{SMDPL}) the signal of the bispectrum of the reference, specially in the case in which non-local information is included in the iterative process. Given that the bispectrum is not calibrated within the procedure, such good agreement between the signals shown in Fig.\ref{Pk_patchy_OldBias} can be regarded as a direct consequence of both the high signal-to-noise in the measurement of the CPD and the inclusion of relevant non-local dependencies such as the cosmic-web classification. Note that in this regard, \citet[][]{2019arXiv190606109B} showed that the agreement seen in Fig.~\ref{Pk_patchy_OldBias} was not present when lower mass resolutions or larger cosmological volumes are used as a reference (using the same \emph{non-local} properties). We shall return to this result in \S~\ref{sec:nl}. 

In Fig.~\ref{densf} we show a slice through the halo density field as obtained from the reference and the two methods under comparison. On large scales the two methods provide a fair description of the halo number density field (as can be inferred from the measurements of power spectrum in Fig.~\ref{Pk_patchy_OldBias}). The evident differences are observed on small scales, where the \pat method tends to break structures such as filaments and sheets and mildly populate voids, all this at the expense of overpopulating high density regions (e.g. knots). In order to quantify this, in Fig.~\ref{biasF} we show the halo-bias in the form of contours of the joint probability distribution, $P(x,y)$ with $x\equiv \log (1+\delta_{\rm m})$ and $y\equiv \log (1+\delta_{h})$. This figure shows that, despite the overall good agreement among the methods, the CPD from the parametric approach shows deviations with respect to the reference, which give rise to the discrepancies observed in the summary statistics showed in Fig.~\ref{Pk_patchy_OldBias}. In order further bring  these differences to light, in Fig.~\ref{bias_pdf} we show examples of conditional probability distribution $P(y|x)$ in three different bins of DM density (or $x$) measured for the full samples (top panels) and in different cosmic web types (from the second to the fourth row). From the first row it is clear how \pat tends to populate cells (or regions) of high density (e.g. $\langle x \rangle = 2.1$) with more halos, as compared to the reference. Approximately the same behavior appears when the CPD is measured in knots and filaments (second and third row, respectively), while similar behaviors are observed for the set of sheets and voids.

In general, the improvement gained with \texttt{BAM} over the \pat method (at the full mass resolution) is evident, suggesting that a parametrisation of the halo-bias as introduced in \S~\ref{sec:patchy} cannot capture the full complexity of the clustering of dark matter haloes at this mass limit. Evidently, a path to close this gap would be that of increasing the dimensionality of the parameter space to be explored. One clear extension would be that of including higher order terms of the dark matter overdensity in Eq.~(\ref{eq:deterministic_nonlocal}). Nevertheless, we consider that given the performance of the non-parametric approach, in which we can capture almost all the physical information of the halo bias without any fitting parameter, makes it worth to put more efforts in pushing the latter towards better performances in view of the construction of mock catalogs for the forthcoming large-scale structure experiments. We are currently making efforts in that direction (F-S Kitaura et al., in preparation) and on the inclusion of velocities (Balaguera et al. in preparation). 
\begin{figure}
\hspace{-0.85cm}
  \includegraphics[width=14cm]{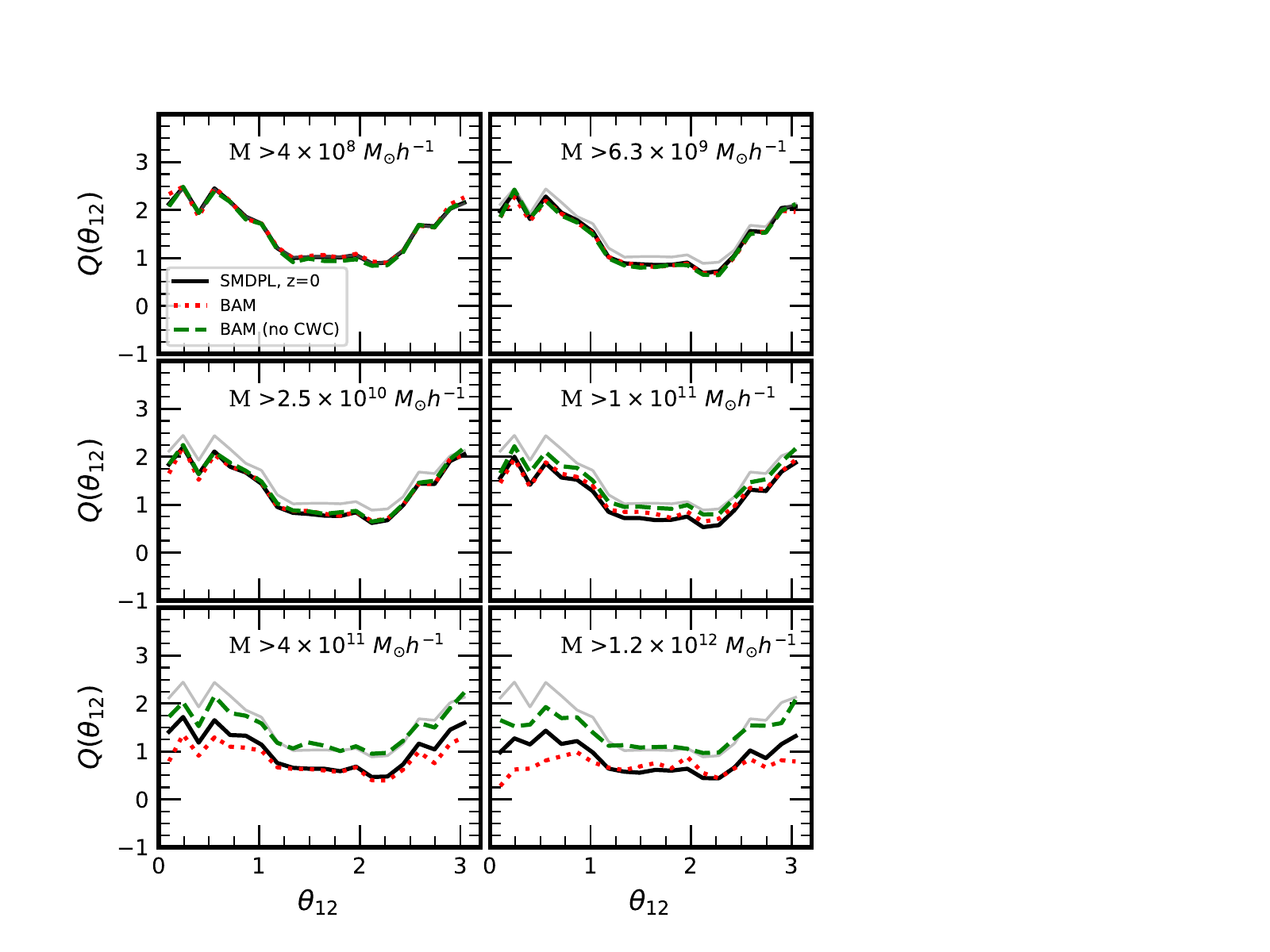}
\vspace{-0.5cm}
\caption{Reduced bispectrum  $Q(\theta_{12})$ measured as a function of the angle between wave-vectors $\theta_{12}=\vec{k}_{1}\cdot \vec{k}_{2}$ (with $|\vec{k}_1|=0.1\,h\, {\rm{Mpc}^{-1}}$ and $|\vec{k}_2|=0.2\,h\, {\rm{Mpc}}^{-1}$), for different halo mass-cuts at $z=0$. Solid lines represents the measurements from the \texttt{SMDPL} simulation. Dotted and dashed lines correspond to the results from \texttt{BAM}, with and without cosmic web classification respectively. In all panels, the grey line represents the measurements from the reference in the  lower mass cut. The grey line represents the bispectrum of the full sample.} 
  \label{bias_bam_masscuts}
 \end{figure}

\section{A signature of non-local bias?}\label{sec:nl}
In \S~\ref{sec:bam} we showed that the non-parametric approach to halo-bias generates an accurate description of the three-point statistics at the mass resolution and volume of the \texttt{SMDPL}  simulation. Furthermore, Fig.~\ref{Pk_patchy_OldBias} demonstrated that the inclusion of non-local properties (viz., cosmic-web classification and the mass of collapsing regions) yields a reduced bispectrum which agrees within $\sim3\%$ ($1\sigma_B$) with that of the reference. Interestingly, this difference is increased to $\sim5\%$ ($2-3\sigma_B$) when only local information is implemented in the bias mapping. We argue that this represents an smoking-gun for the signatures of non-local halo-bias \citep[see e.g][]{2013PhRvD..87h3002S}. 

In view of coming missions such as Euclid or DESI, it is interesting to assess such signature for higher mass cuts, specially those characterizing the selection function of the aforementioned missions \footnote{For instance, Euclid will observe emission line galaxies residing roughly in halos of mass $\sim 10^{11}h^{-1}{\rm M}_\odot$ \citep[see e.g.][]{doi:10.1093/mnras/stx1980}.}. To that aim, we perform the same calibration procedures in \texttt{BAM}, using as a reference the sub-samples of the full \texttt{SMDPL} simulation selected by mass-cuts. Figure \ref{bias_bam_masscuts} shows the results, ranging from a sample with mass $ \sim  4\times  10^{8}{\rm M}_{\odot}h^{-1}$ up to $1.2\times 10^{12}{\rm M}_{\odot}h^{-1}$. \texttt{BAM} replicates the bispectrum of the reference with non-local information (dotted-lines) within $\sim3-6\%$ (or roughly within the error bars from Eq.~(\ref{err_bis})), up to masses of $2.5\times 10^{10}{\rm M}_{\odot}h^{-1}$. If the non-local information is not included (dashed lines), the signal is correct within $\sim6-12\%$ ($\sim 2-3\sigma_B$ as in  Fig.~\ref{Pk_patchy_OldBias}). At $10^{11}{\rm M}_{\odot}h^{-1}$, a systematic bias in the estimates of the bispectrum appears, represented by a $\sim8\%$ ($\sim 1-2\sigma_B$) difference (with non-local dependencies) and $\sim20\%$ ($\sim 3-4\sigma_B$, with only local dependencies). The discrepancy at higher mass-cuts increases although the non-local prediction is able to follow the three-point statistics amplitude of the reference. The reason for this can reside in either the low number count statistics at such high mass-cuts (passing from a Poisson signal-to-noise of $\sim 30$ from the full sample to $\sim 1$ for the highest mass-cut used in Fig.~\ref{bias_bam_masscuts}), or in the lack of other properties of the ADMF taken into account within the \texttt{BAM} procedure. 

\begin{figure}
\hspace{-0.8cm}
  \includegraphics[width=9.6cm]{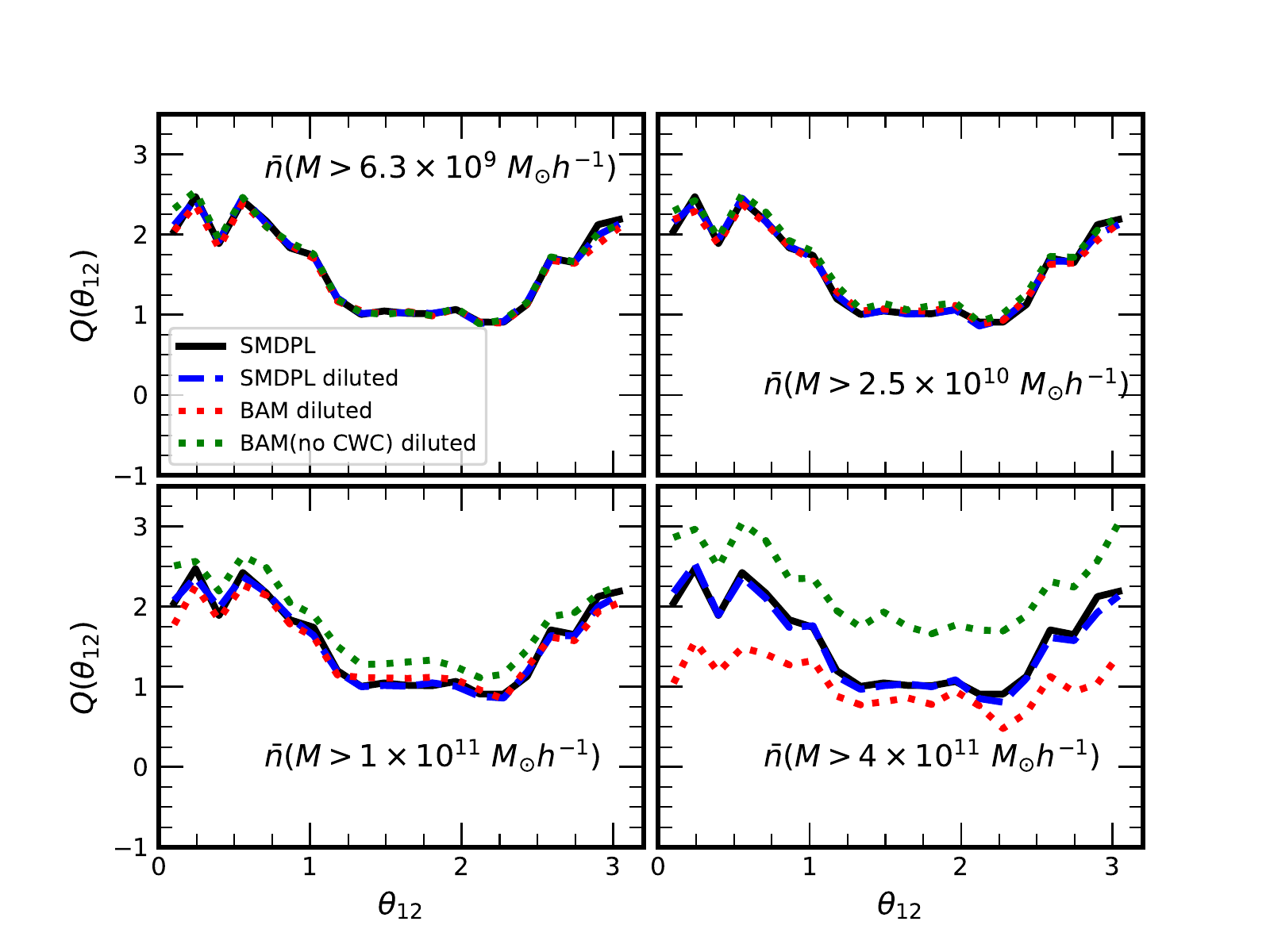}
\vspace{-0.5cm}
\caption{Reduced bispectrum measured from the \texttt{SMDPL} simulation using the full halo catalog (solid line) and diluted (long-dashed line) to different mean number densities characterized by different mass cuts (shown in each panel) (using the same triangle configuration as in Fig.~\ref{bias_bam_masscuts}). The green and red dotted lines represent the bispectrum obtained from \texttt{BAM} calibrated with each diluted versions of the \texttt{SMDPL}, with and without non-local information, respectively.} 
  \label{bias_bam_downsampled}
 \end{figure}
In order to verify whether the behavior captured in Fig.~\ref{bias_bam_masscuts} have physical grounds or, on the contrary are due to a low signal-to-noise number counts, we have performed the following test. We have down-sampled the full \texttt{SMDPL} simulation to different mean number densities (associated to different mass-cuts within the same volume) and verified at which level the reduced bispectrum generated by \texttt{BAM} was still compatible with that of the reference. Note again that, in each case, \texttt{BAM} is calibrated to the down-sampled \texttt{SMDPL} and not the entire \texttt{SMDPL} as in the case of the mass cuts.
The results, shown in Fig.~\ref{bias_bam_downsampled} demonstrates that up to mean number densities associated to mass cuts of $\sim 10^{11}{\rm M}_{\odot}h^{-1}$, \texttt{BAM} does a good job at reproducing the reference bispectrum using non-local information. At that mass-cut, we already discussed that, using only the local information, yields a systematic bias which amounts to $\sim20\%$. 

Thus, this tests concludes that the mean number density associated to this mass-cut marks a threshold up to which the use of non-local information in the measurement of the CPD can reproduce the bispectrum to $<1\sigma_B$ precision. Translating this to Fig.~\ref{bias_bam_masscuts}, we can theorize that the signal at the mass-cut $10^{11}{\rm M}_{\odot}h^{-1}$ is not dominated by the signal-to-noise and hence, the discrepancy (of the order of $1\sigma_B$) between the product of \texttt{BAM} and the reference can be due to lack of physical dependencies, not accounted for in the measurements of the CPD. Including such dependencies is out of the scope of this paper, and will be addressed in a forthcoming publication (F-S. Kitaura et al., in preparation). Note that Fig.~\ref{bias_bam_downsampled} also indicates that we can trust both the local and non-local calibrations for number densities corresponding to masses lower than $\sim 6\times10^{9}{\rm M}_{\odot}h^{-1}$, and thus, the discrepancy of Fig.~\ref{Pk_patchy_OldBias} between the \texttt{BAM} can be considered as a possible signature of non-locality.

\section{Conclusions}
In this paper we have compared the performance of two calibrated methods to construct mock catalogs of dark matter haloes. The methods \pat \citep[][]{2014MNRAS.439L..21K} and \texttt{BAM} \cite[][]{2019MNRAS.483L..58B} represent examples of parametric and non-parametric approaches, respectively, to the idea of mapping the halo-bias onto a dark matter density field in order to reproduce the two-point statistics of a reference simulation. We have shown that a parametric approach, in which a mean halo-bias and the scatter around that mean is modeled with a set of parameters (including more complex versions of the widely used power-law bias), cannot account for the full complexity of the halo statistics at a mass scale of the resolution of the Small MultiDark simulation ($\sim 2\times10^{8}{\rm M}_{\odot}\,h^{-1}$), used as a reference. On the other hand, the non-parametric approach (\texttt{BAM})  generates a good description of the halo distribution up to three-point statistics, thus demonstrating to be properly tailored and suited to capture the main physical properties of the halo bias. As it is clear that larger parameter spaces can be argued in favor of a parametric approach (and perhaps new functional dependencies), we conclude that \texttt{BAM} has the potential to account for high order statistics in a transparent way, given that the physical content can be understood and no tuned parameters are requested.
Furthermore, we have commented on the possibility of detecting non-local bias with \texttt{BAM}, when focused mass scales typical of experiments such as DESI. This will be addressed in a forthcoming publication (F-S Kitaura et al., in preparation).

In summary, we conclude that although parametric approaches to halo-bias mapping are successful at reproducing halo statistics at halo masses of $>10^{12}\,{\rm M}_{\odot}\,h^{-1}$(\citealt{2017MNRAS.472.4144V}), at lower mass scales ($>10^{8}{\rm M}_{\odot}\,h^{-1}$) these models of halo bias (including stochasticity) cannot generate the same precision in the summary statistics of the halo (and sub-halo) distribution. Instead, we have shown that non-parametric approaches (in particular \texttt{BAM}), have the potential to reproduce up to the three point statistics of a halo distribution at low mass scales, where non linear-clustering and non-local dependencies are likely to be dominant.
We note however that provided that our reference sample is composed of halos and sub-halos, the parametric prescriptions are not expected to be accurate, as these were initially motivated for host-haloes. Nonetheless, foreseeing applications for forthcoming galaxy surveys probing emission line galaxies (ELGs), which populate both halos and sub-halos \citep[see e.g.][]{2018MNRAS.475.2530O}, we believe that it is important to show that our methodologies are able to reproduce the clustering of a sample including sub-structures. A clear assessment of the applicability of both methods in such context requires a reliable ELGs reference catalog. Our efforts are currently focused towards that direction.

\section*{Acknowledgements}
MPI acknowledges financial support from the Spanish Ministry of Economy and Competitiveness (MINECO) under the grant AYA2012-39702-C02-01. ABA acknowledges financial support from the Spanish Ministry of Economy and Competitiveness (MINECO) under the Severo Ochoa program SEV-2015-0548. FSK thanks support from the grants RYC2015-18693, SEV-2015-0548 and AYA2017-89891-P. GY acknowledges support  from  MINECO/FEDER under research grant AYA2015-63810-P and MICIU/FEDER PGC2018-094975-C21.
MPI would like to thank Rodrigo Mench\'on and Matteo Zennaro for their useful discussions.  The \texttt{SMD} simulation has been performed on SuperMUC at LRZ in Munich within the pr87yi project . The CosmoSim database
\url{www.cosmosim.org} that  provides access to the \texttt{SMDPL} simulation and
the Rockstar halo catalogues is a service by the Leibniz Institute for
Astrophysics Potsdam (AIP). We acknowledge the support of the European Research Council through grant ERC-StG/716151.


\bibliographystyle{mnras}
\bibliography{refs} 
\end{document}